\documentclass[11pt]{article}
\usepackage{amssymb,amsmath,amsfonts}
\usepackage{graphicx}
\usepackage{graphics}
\usepackage{eepic,epsfig}

\begin{document}

\title{Spinor Casimir effect for concentric spherical shells in the global
monopole spacetime}
\author{E. R. Bezerra de Mello$^{1}$\thanks{%
E-mail: emello@fisica.ufpb.br}\, and A. A. Saharian$^{1,2}$\thanks{%
E-mail: saharyan@server.physdep.r.am} \\
\\
\textit{$^1$Departamento de F\'{\i}sica-CCEN, Universidade Federal da Para%
\'{\i}ba}\\
\textit{58.059-970, J. Pessoa, PB C. Postal 5.008, Brazil}\vspace{0.3cm}\\
\textit{$^2$Department of Physics, Yerevan State University,}\\
\textit{375025 Yerevan, Armenia}}
\maketitle

\begin{abstract}
In this paper we investigate the vacuum polarization effects associated with
a massive fermionic field due to the non-trivial topology of the global
monopole spacetime and boundary conditions imposed on this field.
Specifically we investigate the vacuum expectation values of the
energy-momentum tensor and fermionic condensate admitting that the field
obeys the MIT bag boundary condition on two concentric spherical shells. In
order to develop this analysis, we use the generalized Abel-Plana summation,
which allows to extract from the vacuum expectation values the contribution
coming from a single sphere geometry and to present the second sphere
induced part in terms of exponentially convergent integrals. In the limit of
strong gravitational field corresponding to small values of the parameter
describing the solid angle deficit in global monopole geometry, the
interference part in the expectation values are exponentially suppressed.
The vacuum forces acting on spheres are presented as the sum of self-action
and interaction terms. Due to the surface divergences, the first one is
divergent and needs additional renormalization, while the second one is
finite for all non-zero distances between the spheres. By making use of zeta
function renormalization technique, the total Casimir energy is evaluated in
the region between two spheres. It is shown that the interaction part of the
vacuum energy is negative and the interaction forces between the spheres are
attractive. Asymptotic expressions are derived in various limiting cases. As
a special case we discuss the fermionic vacuum densities for two spherical
shells on background of the Minkowski spacetime.
\end{abstract}

\bigskip

PACS number(s): 03.70.+k, 04.62.+v, 12.39.Ba

\bigskip

\section{Introduction}

It is well known that different types of topological objects may have been
formed in the early universe after Planck time by the vacuum phase
transition \cite{Kibble,Vilenkin}. Depending on the topology of the vacuum
manifold these are domains walls, strings, monopoles and textures. Among
them, cosmic strings and monopoles seem to be the best candidate to be
observed.

A global monopole is a spherical heavy object formed in the phase transition
of a system composed by a self-coupling scalar field iso-triplet $\varphi
^{a}$, whose original global $O(3)$ symmetry is spontaneously broken to $%
U(1) $. The matter fields play the role of an order parameter which outside
the monopole's core acquires a non-vanishing value. The simplified global
monopole was introduced by Sokolov and Starobinsky \cite{Soko77}. The
gravitational effects of of the global monopole were considered in Ref. \cite%
{B-V}, where a solution is presented which describes a global monopole at
large radial distances. The gravitational effects produced by this object
may be approximated by a solid angle deficit in the (3+1)-dimensional
spacetime whose line element is given by
\begin{equation}
ds^{2}=dt^{2}-dr^{2}-\alpha ^{2}r^{2}\left( d\theta ^{2}+\sin ^{2}\theta
d\phi ^{2}\right) \ .  \label{mmetric}
\end{equation}%
Here the parameter $\alpha ^{2}=1-8\pi G\eta ^{2}$ is smaller than unity and
depends on the energy scale $\eta $ where the global symmetry is
spontaneously broken. The energy-momentum tensor associated with this
topological object has a diagonal form and reads: $T_{0}^{0}=T_{1}^{1}=(1-%
\alpha ^{2})/\alpha ^{2}r^{2}$ and $T_{2}^{2}=T_{3}^{3}=0$. The spacetime
defined by the metric tensor above is not flat: the scalar curvature $%
R=2(1-\alpha ^{2})/\alpha ^{2}r^{2}$ and the solid angle is $\Omega =4\pi
\alpha ^{2}$, smaller than the ordinary one, consequently there is a solid
angle deficit given by $\delta \Omega =4\pi (1-\alpha ^{2})$. It is of
interest to note that the effective metric produced in superfluid $^{3}%
\mathrm{He-A}$ by a monopole is described by the line element (\ref{mmetric}%
) with the negative angle deficit, $\alpha >1$, which corresponds to the
negative mass of the topological object \cite{Volo98}. The quasiparticles in
this model are chiral and massless fermions.

The quantum effects due to the point-like global monopole
spacetime on the matter fields have been considered for massless
scalar \cite{M-L} and fermionic \cite{EVN} fields, respectively
(for the heat-kernels on the generalized cone see
\cite{MKS,Dowk99}). In order to develop this analysis, the scalar
respectively spinor Green functions in this background have been
obtained. The influence of the non-zero temperature on these
polarization effects has been considered in \cite{C-E} for scalar
and fermionic fields. Moreover, the calculation of quantum effects
on massless scalar field in a higher dimensional global monopole
spacetime has also been developed in \cite{E}.

Two important ingredients responsible for vacuum polarization effects are
the non-trivial geometry of the spacetime itself and boundary conditions
obeyed by the matter fields. In this direction, the total Casimir energy
associated with massive scalar field inside a spherical region in the global
monopole background have been analyzed in Refs. \cite{MKS,EVN1} by using the
zeta function regularization procedure. Moreover, bosonic Casimir densities
induced by a single shell have been calculated in \cite{A-M} to higher
dimensional global monopole spacetime by making use of the generalized
Abel-Plana summation formula \cite{Sahmat,Sahrev}. More recently, using also
this formalism, a similar analysis for spinor fields has been developed in
\cite{Saha-Mello}. In these publications two distinct situations were
examined: the vacuum averages inside and outside the shell. The generalized
Abel-Plana summation formula is a powerful instrument which allows to
develop the summation over all discrete modes. By using this formula, all
the physical vacuum polarization local quantities can be separated in two
contributions: boundary dependent parts and independent ones. The boundary
independent contributions are similar to previous results obtained in the
literature for scalar and fermionic fields, \cite{M-L,EVN}, using different
approach. These contributions are divergent and consequently in order to
obtain finite and well defined results, one must apply some regularization
procedure. The boundary dependent contributions, on the other hand, are
finite at any strictly interior or exterior points and do not contain
anomalies. Consequently, they do not require any regularization procedure.

Bosonic Casimir densities induced by two concentric spherical shells in the
global monopole spacetime have been analyzed in \cite{Saha03b}. Here we
shall continue in this line of investigation, calculating spinor densities
for two concentric spherical shells. In this context, we shall admit that
the fermionic field obeys the MIT bag boundary conditions on the spherical
shells. Specifically we shall calculate the vacuum expectation value of the
energy-momentum tensor and the vacuum interaction forces between two shells.
Also the total Casimir energy will be considered.

The Casimir effect associated with fermionic fields in a Minkowski
background, was considered in a great number of papers \cite%
{Bend76,Milt80,Milt81,Milt83,Baac83,Blau88,Eliz98,Cogn01} (for reviews and
additional references see \cite{Grib94,Most97,Plun86,Milt02,Bord01}),
considering that the fields obey the MIT bag model condition. As far as we
know, the most of the previous studies were focused on global quantities,
such as the total vacuum energy and stress on the surface. The density of
the fermionic vacuum condensate for a massless spinor field inside the bag
was investigated in Ref. \cite{Milt81} (see also \cite{Milt02}). As to the
Casimir effect it is of physical interest to calculate not only the total
energy but also the local characteristics of the vacuum, such as the
energy-momentum tensor and vacuum condensate. In addition to describing the
physical structure of the quantum field at a given point, the
energy-momentum tensor acts as the source of gravity in the Einstein
equations. It therefore plays an important role in modelling a
self-consistent dynamics involving the gravitational field \cite{Birr82}.
For $\alpha =1$ case, our calculation is a local extension of the previous
contributions on the fermionic Casimir effect for a spherical shell.

This paper is organized as follows: In Section \ref{sec:vevemt} we consider
the vacuum expectation values of the energy-momentum tensor and the
fermionic condensate in the region between two spheres. Section \ref%
{sec:intforce} is devoted to the investigation of the vacuum interaction
forces between the spheres. In Section \ref{sec:Casen}, by using the zeta
function technique, we investigate the total Casimir energy in the region
between two spheres as a sum of the zero-point energies of elementary
oscillators and various limiting cases are discussed. In Section \ref%
{sec:conc} we present our concluding remarks. In Appendix \ref{sec:app1} by
using the generalized Abel-Plana formula we derive a summation formula for
the series over the eigenmodes of the fermionic vacuum in the region between
two spheres.

\section{Vacuum expectation values of the energy-momentum tensor and the
fermionic condensate}

\label{sec:vevemt} In this section we shall consider a massive spinor field
propagating on a point-like global monopole background described by the line
element (\ref{mmetric}). The dynamics of the field is described by the Dirac
equation
\begin{equation}
i\gamma ^{\mu }(\partial _{\mu }+\Gamma _{\mu })\psi -M\psi =0\ ,
\label{Diraceq}
\end{equation}%
where $\gamma ^{\mu }$ are the Dirac matrices defined in such curved
spacetime, and $\Gamma _{\mu }=\gamma _{\nu }\nabla _{\mu }\gamma ^{\nu }/4$
is the spin connection with $\nabla _{\mu }$ being the standard covariant
derivative operator. Here we shall adopt the following representation for
the Dirac matrices\footnote{%
Notice that using this representation the usual anticommutation relation
between the Dirac matrices, $\{\gamma _{\mu },\ \gamma _{\nu }\}=2g_{\mu \nu
}$ is observed.}
\begin{equation}
\gamma ^{0}=\left(
\begin{array}{cc}
1 & 0 \\
0 & -1%
\end{array}%
\right) \ ,\ \gamma ^{k}=\left(
\begin{array}{cc}
0 & \sigma ^{k} \\
-\sigma ^{k} & 0%
\end{array}%
\right) \ ,  \label{Diracmatr}
\end{equation}%
given in terms of the curved space Pauli $2\times 2$ matrices $\sigma ^{k}$:
\begin{eqnarray}
\sigma ^{1} &=&\left(
\begin{array}{cc}
\cos \theta & e^{-i\phi }\sin \theta \\
e^{i\phi }\sin \theta & -\cos \theta%
\end{array}%
\right) \ ,\quad \sigma ^{2}=\frac{1}{\alpha r}\left(
\begin{array}{cc}
-\sin \theta & e^{-i\phi }\cos \theta \\
e^{i\phi }\cos \theta & \sin \theta%
\end{array}%
\right) \ ,  \notag \\
\sigma ^{3} &=&\frac{i}{\alpha r\sin \theta }\left(
\begin{array}{cc}
0 & -e^{-i\phi } \\
e^{i\phi } & 0%
\end{array}%
\right) .  \label{Pauli}
\end{eqnarray}%
These matrices satisfy the relation
\begin{equation}
\sigma ^{l}\sigma ^{k}=\gamma ^{lk}+i\frac{\varepsilon ^{lkm}}{\sqrt{\gamma }%
}\gamma _{mp}\sigma ^{p}\ ,  \label{Pauli1}
\end{equation}%
where $\gamma ^{lk}=-g^{lk}$ are the spatial components of metric tensor and
$\gamma $ is the corresponding determinant. $\varepsilon ^{lkm}$ is the
totally anti-symmetric symbol with $\varepsilon ^{123}=1$, and the latin
indices $i,k,\cdots $ run over values $1,\ 2,\ 3$.

In this paper we are interested in the vacuum expectation value (VEV) of the
energy-momentum tensor in the region between two spherical shells concentric
with the global monopole on which the field obeys bag boundary conditions:
\begin{equation}
\left( 1+i\gamma ^{\mu }n_{\mu }^{(w)}\right) \psi \big| _{r=w}=0\ ,\;w=a,b,
\label{boundcond1}
\end{equation}%
where $a$ and $b$ are the radii for the spheres, $a<b$, $n_{\mu
}^{(w)}=n^{(w)}\delta _{\mu }^{1}$ is the outward-pointing normal to the
boundaries. Here and below we use the notations $n^{(a)}=-1$, $n^{(b)}=1$.
To find the VEV for the energy-momentum tensor we expand the field operator
in terms of the complete set of single-particle states $\left\{ \psi _{\chi
}^{(+)},\psi _{\chi }^{(-)}\right\} $:
\begin{equation}
\hat{\psi}=\sum_{\chi }\left( \hat{a}_{\chi }\psi _{\chi }^{(+)}+\hat{b}%
_{\chi }^{+}\psi _{\chi }^{(-)}\right) ,  \label{operatorexp}
\end{equation}%
where $\hat{a}_{\chi }$ is the annihilation operator for particles, and $%
\hat{b}_{\chi }^{+}$ is the creation operator for antiparticles. Now we
substitute the expansion (\ref{operatorexp}) and the analog expansion for
the operator $\hat{\bar{\psi}}$ into the corresponding expression of the
energy-momentum tensor for the spinor fields,
\begin{equation}
T_{\mu \nu }\left\{ \hat{\bar{\psi}},\hat{\psi}\right\} =\frac{i}{2}\left[
\hat{\bar{\psi}}\gamma _{(\mu }\nabla _{\nu )}\hat{\psi}-(\nabla _{(\mu }%
\hat{\bar{\psi}})\gamma _{\nu )}\hat{\psi}\right] \ .  \label{EMTform}
\end{equation}%
By making use the standard anticommutation relations for the annihilation
and creation operators, for the VEVs one finds the following mode-sum
formula
\begin{equation}
\left\langle 0\left\vert T_{\mu \nu }\right\vert 0\right\rangle =\sum_{\chi
}T_{\mu \nu }\left\{ \bar{\psi}_{\chi }^{(-)}(x),\psi _{\chi
}^{(-)}(x)\right\} \ ,  \label{modesum}
\end{equation}%
where $|0\rangle $ is the amplitude for the corresponding vacuum. For the
problem under consideration the eigenfunctions have the form \cite%
{Saha-Mello}
\begin{eqnarray}
\psi _{\chi }^{(\pm )} &=&\frac{Ae^{\mp iEt}}{\sqrt{r}}\left(
\begin{array}{c}
Z_{\nu _{\sigma }}(kr)\Omega _{jlm} \\
in_{\sigma }Z_{\nu _{\sigma }+n_{\sigma }}(kr)\frac{k(\hat{n}\cdot \vec{%
\sigma})}{\pm E+M}\Omega _{jlm}%
\end{array}%
\right) ,  \label{eigfunc} \\
E &=&\sqrt{k^{2}+M^{2}},\;l=j-\frac{n_{\sigma }}{2},\;n_{\sigma
}=(-1)^{\sigma },
\end{eqnarray}%
where $\vec{\sigma}=(\sigma ^{1},\sigma ^{2},\sigma ^{2})$, and $\hat{n}=%
\vec{r}/r$. These functions are specified by the set of quantum numbers $%
\chi =(\sigma kjm)$, where $0\leq k<\infty $, $j=1/2,3/2,\ldots $ denotes
the value of the total angular momentum, $m=-j,\ldots ,j$ determines the
value for its projection, $\sigma =0,1$ specifies two types of
eigenfunctions with different parities corresponding to $l=j-n_{\sigma }/2$
with $l$ being the orbital momentum.\ In formula (\ref{eigfunc}), $\Omega
_{jlm}$ are the standard spinor spherical harmonics whose explicit form is
given in Ref. \cite{Berest}, and $Z_{\nu _{\sigma }}(x)$ represents the
cylindrical Bessel function of the order
\begin{equation}
\nu _{\sigma }=\frac{j+1/2}{\alpha }-\frac{n_{\sigma }}{2}.
\label{nuplusmin}
\end{equation}%
The functions (\ref{eigfunc}) are orthonormalized by the condition
\begin{equation}
\int d^{3}x\sqrt{\gamma }\psi _{\chi }^{(\eta )+}\psi _{\chi ^{\prime
}}^{(\eta ^{\prime })}=\delta _{\eta \eta ^{\prime }}\delta _{\chi \chi
^{\prime }}\ ,\ \eta \ ,\eta ^{\prime }=\pm ,  \label{normcond}
\end{equation}%
from which the normalization constant $A$ can be determined. As in mode-sum
formula (\ref{modesum}) the negative frequency modes are employed, in the
discussion below we shall consider the eigenfunctions (\ref{eigfunc}) with
the lower sign. Note that in terms of the function $Z_{\nu _{\sigma }}(kr)$
the boundary conditions (\ref{boundcond1}) take the form%
\begin{equation}
Z_{\nu _{\sigma }}(kw)=-Z_{\nu _{\sigma }+n_{\sigma }}(kw)\frac{%
n^{(w)}n_{\sigma }k}{\sqrt{k^{2}+M^{2}}-M},  \label{boundcondZ}
\end{equation}%
with $w=a,b$.

In the region between two spherical shells concentric with the global
monopole, the function $Z_{\nu _{\sigma }}(kr)$ is a linear combination of
the Bessel, $J_{\nu }(z)$, and Neuamnn, $Y_{\nu }(z)$, functions. The
coefficient in this linear combination is determined from the boundary
condition (\ref{boundcond1}) on the sphere $r=a$ and one obtains%
\begin{equation}
Z_{\nu _{\sigma }}^{(a)}(kr)=g_{\nu _{\sigma }}^{(a)}(ka,kr)\equiv J_{\nu
_{\sigma }}(kr)\tilde{Y}_{\nu _{\sigma }}^{(a)}(ka)-Y_{\nu _{\sigma }}(kr)%
\tilde{J}_{\nu _{\sigma }}^{(a)}(ka),  \label{gnukakr}
\end{equation}%
where for a given function $F(z)$ we use the notation
\begin{equation}
\tilde{F}^{(w)}(z)\equiv zF^{\prime }(z)+\left[ n^{(w)}\left( \mu _{w}-\sqrt{%
z^{2}+\mu _{w}^{2}}\right) -n_{\sigma }\nu _{\sigma }\right] F(z)\ ,
\label{tildenot}
\end{equation}%
with $w=a,b$, and $\mu _{w}=Mw$. Now from the boundary condition on the
outer sphere one finds that the eigenvalues for $k$ are solutions to the
equation%
\begin{equation}
C_{\nu _{\sigma }}(b/a,ka)\equiv \tilde{J}_{\nu _{\sigma }}^{(a)}(ka)\tilde{Y%
}_{\nu _{\sigma }}^{(b)}(kb)-\tilde{Y}_{\nu _{\sigma }}^{(a)}(ka)\tilde{J}%
_{\nu _{\sigma }}^{(b)}(kb)=0.  \label{Cnu}
\end{equation}%
Below we denote by $\gamma _{\nu _{\sigma },s}=ka$, $s=1,2,\ldots ,$ the
positive roots to equation (\ref{Cnu}), arranged in ascending order, $\gamma
_{\nu _{\sigma },s}<\gamma _{\nu _{\sigma },s+1}$. Substituting the
eigenfunctions into the normalization integral (\ref{normcond}) with the
integration over the region $a\leqslant r\leqslant b$ and using the standard
integrals for cylindrical functions (see, for instance, \cite{Prud86}), for
the normalization coefficient of the negative frequency modes one finds%
\begin{equation}
A^{2}=\frac{\pi ^{2}k(\sqrt{k^{2}+M^{2}}-M)}{8\alpha ^{2}a\sqrt{k^{2}+M^{2}}}%
T_{\nu }(\eta ,ka),\;ka=\gamma _{\nu _{\sigma },s},  \label{Acoef}
\end{equation}%
where $T_{\nu }(\eta ,ka)$ is defined in Appendix \ref{sec:app1}.

Since the bulk and boundary geometries are spherically symmetric and static,
the vacuum energy-momentum tensor has diagonal form, moreover $\langle
T_{\theta }^{\theta }\rangle =\langle T_{\phi }^{\phi }\rangle $. So in this
case we can write:
\begin{equation}
\left\langle 0\left\vert T_{\mu }^{\nu }\right\vert 0\right\rangle =\mathrm{%
diag}(\varepsilon ,-p,-p_{\perp },-p_{\perp })\ ,  \label{diagform}
\end{equation}%
with the energy density $\varepsilon $, radial, $p$, and azimuthal, $%
p_{\perp }$, pressures. As a consequence of the continuity equation $\nabla
_{\nu }\left\langle 0\left\vert T_{\mu }^{\nu }\right\vert 0\right\rangle =0$%
, these functions are related by the equation
\begin{equation}
r\frac{dp}{dr}+2(p-p_{\perp })=0\ ,  \label{conteq}
\end{equation}%
which means that the $r$-dependence of the radial pressure necessarily leads
to the anisotropy in the vacuum stresses.

Substituting eigenfunctions (\ref{eigfunc}) into Eq. (\ref{modesum}), the
summation over the quantum number $m$ can be done by using standard
summation formula for the spherical harmonics. For the energy-momentum
tensor components one finds
\begin{equation}
q(r)=\frac{-\pi }{32\alpha ^{2}a^{3}r}\sum_{j=1/2}^{\infty
}(2j+1)\sum_{\sigma =0,1}\sum_{s=1}^{\infty }T_{\nu _{\sigma }}(\eta
,z)f_{\sigma \nu _{\sigma }}^{(q)}\left[ z,g_{\nu _{\sigma }}^{(a)}(z,zr/a)%
\right] _{z=\gamma _{\nu _{\sigma },s}},  \label{qrin}
\end{equation}%
with $q=\varepsilon ,\ p,\ p_{\perp }$ and we have introduced the notations
\begin{eqnarray}
f_{\sigma \nu }^{(\varepsilon )}\left[ z,g_{\nu }^{(a)}(z,y)\right] &=&z%
\left[ (\sqrt{z^{2}+\mu _{a}^{2}}-\mu _{a})g_{\nu }^{(a)2}(z,y)+(\sqrt{%
z^{2}+\mu _{a}^{2}}+\mu _{a})g_{\nu +n_{\sigma }}^{(a)2}(z,y)\right] \ ,
\label{fnueps} \\
f_{\sigma \nu }^{(p)}\left[ z,g_{\nu }^{(a)}(z,y)\right] &=&\frac{z^{3}}{%
\sqrt{z^{2}+\mu _{a}^{2}}}\Big[ g_{\nu }^{(a)2}(z,y)+g_{\nu +n_{\sigma
}}^{(a)2}(z,y)  \notag \\
&& -\frac{2\nu +n_{\sigma }}{y}g_{\nu }^{(a)}(z,y)g_{\nu +n_{\sigma
}}^{(a)}(z,y)\Big] \ ,  \label{fnup} \\
f_{\sigma \nu }^{(p_{\perp })}\left[ z,g_{\nu }^{(a)}(z,y)\right] &=&\frac{%
z^{3}(2\nu +n_{\sigma })}{2y\sqrt{z^{2}+\mu _{a}^{2}}}g_{\nu
}^{(a)}(z,y)g_{\nu +n_{\sigma }}^{(a)}(z,y)\ ,  \label{fnupperp}
\end{eqnarray}%
with $g_{\nu +n_{\sigma }}^{(a)}(z,y)\equiv J_{\nu _{+n_{\sigma }}}(y)\tilde{%
Y}_{\nu }^{(a)}(z)-Y_{\nu _{+n_{\sigma }}}(y)\tilde{J}_{\nu }^{(a)}(z)$. The
vacuum expectation values given by formulae (\ref{qrin}) are divergent and
need some regularization procedure. To make them finite we can introduce a
cutoff function $\Phi _{\lambda }(z)$, $z=\gamma _{\nu _{\sigma },s}$ with
the cutoff parameter $\lambda $, which decreases sufficiently fast with
increasing $z$ and satisfies the condition $\Phi _{\lambda }\to  1$, $%
\lambda \to  0$. Now we apply to the sum over $s$ the summation
formula derived in Appendix \ref{sec:app1}. As a function $h(z)$
in this formula we take $h(z)=f_{\sigma \nu _{\sigma
}}^{(q)}\left[ z,g_{\nu _{\sigma }}^{(a)}(z,zr/a)\right] \Phi
_{\lambda }(z)$. As it has been pointed out in Appendix
\ref{sec:app1}, the function $f_{\sigma \nu _{\sigma
}}^{(q)}\left[ z,g_{\nu _{\sigma }}^{(a)}(z,zr/a)\right] $
satisfies relation (\ref{relforf}) and, hence, the part of the
integral on the right of formula (\ref{cor3form}) over the
interval $(0,\mu _{a})$ vanishes after removing the cutoff. As a
result the components of the vacuum
energy-momentum tensor can be presented in the form%
\begin{equation}
q(r)=q_{1a}(r)+q_{ab}(r),  \label{qr1}
\end{equation}%
where
\begin{equation}
q_{1a}(r)=\frac{-1}{8\pi \alpha ^{2}a^{3}r}\sum_{j=1/2}^{\infty
}(2j+1)\sum_{\sigma =0,1}\int_{0}^{\infty }dx\frac{f_{\sigma \nu _{\sigma
}}^{(q)}\left[ x,g_{\nu _{\sigma }}^{(a)}(x,xr/a)\right] }{\tilde{J}_{\nu
_{\sigma }}^{(a)2}(x)+\tilde{Y}_{\nu _{\sigma }}^{(a)2}(x)},  \label{q1out}
\end{equation}%
and
\begin{equation}
q_{ab}(r)=\frac{-1}{8\pi ^{2}\alpha ^{2}r}\sum_{j=1/2}^{\infty
}(2j+1)\sum_{\sigma =0,1}\sum_{\beta =\pm }\int_{M}^{\infty }dx\Omega _{a\nu
_{\sigma }}^{(\beta )}(ax,bx)F_{\sigma \nu _{\sigma }}^{(q\beta )}\left[
ax,G_{\nu _{\sigma }}^{(a\beta )}(ax,rx)\right] ,  \label{qabr}
\end{equation}%
with $\Omega _{a\nu _{\sigma }}^{(\beta )}(ax,bx)$ defined in Appendix \ref%
{sec:app1}. In formula (\ref{qabr}) we have introduced the notations%
\begin{eqnarray}
F_{\sigma \nu }^{(\varepsilon \beta )}\left[ x,G_{\nu }^{(a\beta )}(ax,y)%
\right] &=&x\left[ (\sqrt{x^{2}-M^{2}}+\beta iM)G_{\nu }^{(a\beta
)2}(ax,y)\right.  \notag \\
&&\left. -(\sqrt{x^{2}-M^{2}}-\beta iM)G_{\nu +n_{\sigma }}^{(a\beta
)2}(ax,y)\right]  \label{Feps} \\
F_{\sigma \nu }^{(p\beta )}\left[ x,G_{\nu }^{(a\beta )}(ax,y)\right] &=&%
\frac{x^{3}}{\sqrt{x^{2}-M^{2}}}\left[ G_{\nu }^{(a\beta )2}(ax,y)-G_{\nu
+n_{\sigma }}^{(a\beta )2}(ax,y)\right.  \notag \\
&&\left. -n_{\sigma }\frac{2\nu +n_{\sigma }}{y}G_{\nu }^{(a\beta
)}(ax,y)G_{\nu +n_{\sigma }}^{(a\beta )}(ax,y)\right]  \label{Fp} \\
F_{\sigma \nu }^{(p_{\perp }\beta )}\left[ x,G_{\nu }^{(a\beta )}(ax,y)%
\right] &=&\frac{x^{3}(2\nu +n_{\sigma })n_{\sigma }}{2y\sqrt{x^{2}-M^{2}}}%
G_{\nu }^{(a\beta )}(ax,y)G_{\nu +n_{\sigma }}^{(a\beta )}(ax,y)\ ,
\label{Fpperp}
\end{eqnarray}%
where
\begin{eqnarray}
G_{\nu }^{(w\pm )}(x,y) &=&I_{\nu }(y)K_{\nu }^{(w\pm )}(x)-K_{\nu
}(y)I_{\nu }^{(w\pm )}(x),\;w=a,b,  \label{Gnu} \\
G_{\nu +n_{\sigma }}^{(w\pm )}(x,y) &=&I_{\nu +n_{\sigma }}(y)K_{\nu
}^{(w\pm )}(x)+K_{\nu +n_{\sigma }}(y)I_{\nu }^{(w\pm )}(x).  \label{Gnunsig}
\end{eqnarray}%
In these formulae and below for a given function $F(z)$ we use the notation $%
F^{(w\pm )}(z)$ defined by formula (\ref{Fbarpm}) in Appendix \ref{sec:app1}%
. The expression on the right of Eq. (\ref{qabr}) can be further simplified
by taking into account the relations%
\begin{eqnarray}
\frac{I_{\nu _{0}}^{(w\pm )}(x)}{K_{\nu _{0}}^{(w\pm )}(x)} &=&-\frac{I_{\nu
_{1}}^{(w\mp )}(x)}{K_{\nu _{1}}^{(w\mp )}(x)},\quad  \label{relsig01} \\
\frac{G_{\nu _{\sigma }+n_{\sigma }}^{(w\pm )}(x,y)}{K_{\nu _{\sigma
}}^{(w\pm )}(x)} &=&\frac{G_{\nu _{\sigma ^{\prime }}}^{(w\mp )}(x,y)}{%
K_{\nu _{\sigma ^{\prime }}}^{(w\mp )}(x)},  \label{relsig012}
\end{eqnarray}%
where $\sigma ^{\prime }=0$ for $\sigma =1$ and $\sigma ^{\prime }=1$ for $%
\sigma =0$. From these relations and formulae (\ref{Feps})-(\ref{Fpperp}), (%
\ref{Oma}) it follows that%
\begin{eqnarray}
K_{\nu _{0}}^{(a\pm )2}(x)\Omega _{a\nu _{0}}^{(\pm )}(x,y) &=&-K_{\nu
_{1}}^{(a\mp )2}(x)\Omega _{a\nu _{1}}^{(\mp )}(x,y),  \label{relsig02} \\
\frac{F_{0\nu _{0}}^{(q\pm )}\left[ x,G_{\nu _{0}}^{(a\pm )}(x,y)\right] }{%
K_{\nu _{0}}^{(a\pm )2}(x)} &=&-\frac{F_{1\nu _{1}}^{(q\mp )}\left[ x,G_{\nu
_{1}}^{(a\mp )}(x,y)\right] }{K_{\nu _{1}}^{(a\mp )2}(x)}.  \label{relsog021}
\end{eqnarray}%
By using these formulae we see that the different parities corresponding to $%
\sigma =0,1$ give the same contributions to the VEVs of the energy-momentum
tensor and the formula (\ref{qabr}) can be written as%
\begin{equation}
q_{ab}(r)=\frac{-1}{2\pi ^{2}\alpha ^{2}r}\sum_{l=1}^{\infty }l\sum_{\beta
=\pm }\int_{M}^{\infty }dx\Omega _{a\nu }^{(\beta )}(ax,bx)F_{1\nu
}^{(q\beta )}\left[ ax,G_{\nu }^{(a\beta )}(ax,rx)\right] ,  \label{qabrl}
\end{equation}%
where
\begin{equation}
\nu \equiv \nu _{1}=\frac{l}{\alpha }+\frac{1}{2}\ ,  \label{notnunu1}
\end{equation}%
and the notation (\ref{Fbarpm}) is specified to%
\begin{equation}
F^{(w\pm )}(z)=zF^{\prime }(z)+\left( \nu +n^{(w)}\mu _{w}\mp in^{(w)}\sqrt{%
z^{2}-\mu _{w}^{2}}\right) F(z)\ .  \label{Fwpm}
\end{equation}

As it has been shown in Ref. \cite{Saha-Mello}, the term (\ref{qr1})
presents the components for the vacuum energy-momentum tensor in the case of
a single spherical shell with radius $a$ in the region $r>a$. After the
subtraction of the part coming from the global monopole geometry without
boundaries, $q_{m}(r)$, the components $q_{1a}(r)$ are presented in the form%
\begin{equation}
q_{1a}(r)=q_{m}(r)+q_{a}(r),  \label{q1r}
\end{equation}%
where the part%
\begin{equation}
q_{a}(r)=\frac{-1}{2\pi ^{2}\alpha ^{2}r}\sum_{l=1}^{\infty }l\sum_{\beta
=\pm }\int_{M}^{\infty }dx\frac{I_{\nu }^{(a\beta )}(ax)}{K_{\nu }^{(a\beta
)}(ax)}F_{1\nu }^{(q\beta )}\left[ x,K_{\nu }(rx)\right] \ ,  \label{qar}
\end{equation}%
is induced by a single sphere with radius $a$ in the region $r>a$. This
quantity diverges on the boundary $r=a$ with the leading divergence $%
(r-a)^{-3}$ for the energy density and the azimuthal stress, and $(r-a)^{-2}$
for the radial stress (see Ref. \cite{Saha-Mello}).

Note that by using the identities%
\begin{eqnarray}
&&\frac{I_{\nu }^{(a\beta )}(ax)}{K_{\nu }^{(a\beta )}(ax)}F_{1\nu
}^{(q\beta )}\left[ x,K_{\nu }(rx)\right] =\frac{K_{\nu }^{(b\beta )}(bx)}{%
I_{\nu }^{(b\beta )}(bx)}F_{1\nu }^{(q\beta )}\left[ x,I_{\nu }(rx)\right]
\notag \\
&&+\sum_{w=a,b}n^{(w)}\Omega _{w\nu }^{(\beta )}(ax,bx)F_{1\nu }^{(q\beta )}
\left[ x,G_{\nu }^{(w\beta )}(wx,rx)\right] ,  \label{ident2}
\end{eqnarray}%
with the notation
\begin{equation}
\Omega _{b\nu }^{(\beta )}(ax,bx)=\frac{I_{\nu }^{(a\beta )}(ax)/I_{\nu
}^{(b\beta )}(bx)}{K_{\nu }^{(a\beta )}(ax)I_{\nu }^{(b\beta )}(bx)-I_{\nu
}^{(a\beta )}(ax)K_{\nu }^{(b\beta )}(bx)},  \label{Omb}
\end{equation}%
the vacuum energy-momentum tensor in the region between the spheres can also
be presented in the form%
\begin{equation}
q(r)=q_{m}(r)+q_{b}(r)+q_{ba}(r),  \label{qr2}
\end{equation}%
where%
\begin{equation}
q_{ba}(r)=\frac{-1}{2\pi ^{2}\alpha ^{2}r}\sum_{l=1}^{\infty }l\sum_{\beta
=\pm }\int_{M}^{\infty }dx\Omega _{b\nu }^{(\beta )}(ax,bx)F_{1\nu
}^{(q\beta )}\left[ x,G_{\nu }^{(b\beta )}(bx,rx)\right] ,  \label{qbar}
\end{equation}%
and the quantities $q_{b}(r)$ are the vacuum densities induced by a single
shell with radius $b$ in the region $r<b$. The formula for the latter is
obtained from (\ref{qar}) by the replacements $a\to  b$, $%
I\rightleftarrows K$. The surface divergences in vacuum expectation values
of the energy-momentum tensor are the same as those for a single sphere when
the second sphere is absent and are investigated in Ref. \cite{Saha-Mello}.
In particular, the term $q_{ab}(r)$ ($q_{ba}(r)$) is finite on the outer
(inner) sphere. It follows from here that if we decompose the VEVs as
\begin{equation}
q(r)=q_{m}(r)+\sum_{w=a,b}q_{w}(r)+q^{(ab)}(r),  \label{qdec}
\end{equation}%
when the interference term $q^{(ab)}(r)$ is finite for all values $%
a\leqslant r\leqslant b$. Two equivalent representations for this term are
obtained from formulae (\ref{qabrl}), (\ref{qar}), (\ref{qbar}). It can be
explicitly checked that separate terms on the right of formula (\ref{qdec})
satisfy the continuity equation (\ref{conteq}).

Having the components of the energy-momentum tensor we can find the
corresponding fermionic condensate $\langle 0|\bar{\psi}\psi |0\rangle $
making use of the formula for the trace of the energy-momentum tensor, $%
T_{\mu }^{\mu }=M\bar{\psi}\psi $. It is presented in two equivalent forms
corresponding to $w=a,b$:
\begin{eqnarray}
\langle 0|\bar{\psi}\psi |0\rangle &=&\langle 0_{m}|\bar{\psi}\psi
|0_{m}\rangle +\langle \bar{\psi}\psi \rangle _{w}-\frac{1}{2\pi ^{2}\alpha
^{2}r}\sum_{l=1}^{\infty }l\sum_{\beta =\pm }\int_{M}^{\infty }dx\ \frac{%
x\Omega _{w\nu }^{(\beta )}(ax,bx)}{\sqrt{x^{2}-M^{2}}}  \notag \\
&&\times \left[ (\beta i\sqrt{x^{2}-M^{2}}-M)G_{\nu }^{(w\beta
)2}(wx,rx)\right.  \notag \\
&&+\left. (\beta i\sqrt{x^{2}-M^{2}}+M)G_{\nu -1}^{(w\beta )2}(wx,rx)\right]
,  \label{condens}
\end{eqnarray}%
where the boundary-free part $\langle 0_{m}|\bar{\psi}\psi
|0_{m}\rangle $ and the single sphere-induced part $\langle
\bar{\psi}\psi \rangle _{w}$ are investigated in Ref.
\cite{Saha-Mello}. Alternatively one could obtain formula
(\ref{condens}) applying the summation formula (\ref{cor3form}) to
the corresponding mode-sum $\sum_{\chi }\bar{\psi}_{\chi
}^{(-)}\psi _{\chi }^{(-)}$ for the fermionic condensate. In the
limit $a\to  0$ with fixed values for the other parameters, the
main contribution into the interference parts of the
energy-momentum tensor and the vacuum condensate comes from $l=1$
terms and these quantities vanish as $a^{2/\alpha }$. In the limit
$b\to  \infty $ for a massless field the interference parts
behave like $b^{-2/\alpha -2}$. For a massive field and under the condition $%
Mb\gg 1$ the main contribution comes from the lower limit of the integral
and the VEVs are exponentially suppressed.

The limit $\alpha \ll 1$ corresponds to a large solid angle deficit. In this
limit and for a fixed $r$ the scalar curvature is positive and large. In
accordance with formula (\ref{notnunu1}) the order of the Bessel modified
functions in the formulae for the interference parts is large. By using the
uniform asymptotic expansions for these functions (see, for instance, \cite%
{Abra64}), we can see that the interference parts are suppressed by the
factor $\exp [(2/\alpha )\ln (a/b)]$. As it has been mentioned in
Introduction, the global monopole geometry with a negative solid angle
deficit corresponding to $\alpha >1$ arises in superfluid $^{3}\mathrm{He-A}$%
. So it is of interest to consider also the limit $\alpha \gg 1$.
Note that in the limit $\alpha \to  \infty $ one has $\nu \to
1/2$ and the series over $l$ in the expressions for the VEVs
diverge. It follows from here that for $\alpha \gg 1$ the main
contribution into these series comes from large values $l$ and to
the leading order we can replace the summation
over $l$ by the integration:%
\begin{equation}
\sum_{l=1}^{\infty }lf(\nu )\to  \alpha ^{2}\int_{1/2}^{\infty
}dy\,(y-1/2)f(y),  \label{sumltoint}
\end{equation}%
where we have used the expression for $\nu $ from (\ref{notnunu1}). Now by
taking into account formulae (\ref{qabrl}), (\ref{qbar}), (\ref{condens}),
we conclude that in the limit under consideration the boundary induced VEVs
for the local observables tend to the finite limiting value obtained from
these formulae by the replacement (\ref{sumltoint}). Note that in this case
the scalar curvature for the background spacetime also takes the finite
limiting value $-2/r^{2}$.

In the formulae above taking $\alpha =1$ we obtain the corresponding
quantities for a spinor field in the Minkowski bulk. In this case $\nu
=l+1/2 $ and the Bessel modified functions are expressed in terms of
elementary functions. Note that the previous investigations on the spinor
Casimir effect for a spherical boundary consider mainly global quantities,
such as total vacuum energy or the vacuum forces acting on the boundary. For
the case of a massless spinor the density of the fermionic condensate $%
\langle \bar{\psi}\psi \rangle _{b}$ is investigated in \cite{Milt81} (see
also \cite{Milt02}).

\section{ Vacuum interaction forces between the spheres}

\label{sec:intforce}

An interesting property of the Casimir effect has always been the geometry
dependence of the vacuum forces acting on the boundaries. In this section we
investigate the interaction forces between the spheres as functions on
sphere radii and the solid angle deficit. The vacuum force per unit surface
on the sphere at $r=w$ is determined by the radial stress evaluated at this
point. The corresponding effective pressures can be presented as a sum of
two terms:
\begin{equation}
p^{(w)}=p_{1}^{(w)}+p_{\mathrm{(int)}}^{(w)},\quad w=a,b,  \label{FintD}
\end{equation}%
where the first term on the right is the pressure for a single sphere at $%
r=w $ when the second sphere is absent and the term $p_{\mathrm{(int)}%
}^{(w)} $ is induced by the presence of the second sphere. The term $%
p_{1}^{(w)}$ directly evaluated from the corresponding bulk stress
in the limit $r\to  w$ is divergent due to the surface divergences
in the subtracted VEVs and needs additional renormalization. This
can be done, for example, by applying the generalized zeta
function technique to the corresponding mode-sum. This calculation
lies on the same line with the evaluation of the total Casimir
energy for a single sphere in the global monopole spacetime and
will be presented in the forthcoming paper. The second term on the
right of Eq. (\ref{FintD}) determines the force by which the
fermionic vacuum acts on the sphere due to the modification of the
spectrum for the zero-point fluctuations by the presence of the
second sphere and can be termed as an interaction force. This term
is finite for all nonzero distances between the spheres and is not
affected by the renormalization procedure. As the vacuum
properties are $r$-dependent, there is no a priori reason for the
interaction terms (and also for the total pressures $p^{(w)}$) to
be equal for $w=a$ and $\ w=b$, and the corresponding forces in
general are different. For the sphere at $r=w$, the
interaction term is due to the summand $p^{(ab)}$ for the inner sphere and $%
p^{(ba)}$ for the outer sphere. Substituting into these terms $r=a$ and $r=b$
respectively and using the relations%
\begin{equation}
G_{\nu }^{(w\beta )}(wx,wx)=-1,\;G_{\nu -1}^{(w\beta
)}(wx,wx)=n^{(w)}B^{(\beta )}/x,  \label{Gnuxx}
\end{equation}%
with the notation%
\begin{equation}
B^{(\pm )}=M\mp i\sqrt{x^{2}-M^{2}},  \label{Bnuw}
\end{equation}%
one finds%
\begin{equation}
p_{\mathrm{(int)}}^{(w)}=\frac{1}{2\pi ^{2}\alpha ^{2}w}\sum_{l=1}^{\infty
}l\sum_{\beta =\pm }\int_{M}^{\infty }dx\frac{x\Omega _{w\nu }^{(\beta
)}(ax,bx)}{\sqrt{x^{2}-M^{2}}}\left[ B^{(\beta )2}+\frac{2n^{(w)}l}{w\alpha }%
B^{(\beta )}-x^{2}\right] .  \label{pwint}
\end{equation}%
Now by making use of the Wronskian for the Bessel modified functions it can
be seen that the following relation takes place%
\begin{equation}
\frac{n^{(w)}}{w}\frac{\partial }{\partial w}\ln \left[ 1-\frac{I_{\nu
}^{(a\beta )}(ax)K_{\nu }^{(b\beta )}(bx)}{K_{\nu }^{(a\beta )}(ax)I_{\nu
}^{(b\beta )}(bx)}\right] =\left[ B^{(\beta )2}+\frac{2\nu -1}{w}%
n^{(w)}B^{(\beta )}-x^{2}\right] \Omega _{w\nu }^{(\beta )}(ax,bx).
\label{relderiv1}
\end{equation}%
This allows us to present the vacuum interaction forces in the form%
\begin{equation}
p_{\mathrm{(int)}}^{(w)}=\frac{n^{(w)}}{\pi ^{2}\alpha ^{2}w^{2}}\frac{%
\partial }{\partial w}\sum_{l=1}^{\infty }l\int_{M}^{\infty }dx\frac{x}{%
\sqrt{x^{2}-M^{2}}}\ln \left\vert 1-\frac{I_{\nu }^{(a+)}(ax)K_{\nu
}^{(b+)}(bx)}{K_{\nu }^{(a+)}(ax)I_{\nu }^{(b+)}(bx)}\right\vert ,
\label{pwint1}
\end{equation}%
where we have taken into account that $F^{(w-)}(wx)=F^{(w+)\ast }(wx)$. In
the next section we will show that these forces can also be obtained from
the Casimir energy by standard differentiation with respect to the sphere
radii. In particular, we shall see that in the limit $a,b\to  \infty $%
, with fixed $b-a$, from (\ref{pwint1}) the result is obtained for two
parallel plates with bag boundary conditions in the Minkowski background.
Other asymptotic expansions for the interaction forces can be obtained from
the corresponding formulae for the interaction energy and will be discussed
in the next section.

\section{Casimir energy}

\label{sec:Casen}

Due to the surface divergences in the VEVs of the local physical
observables, the renormalized total vacuum energy can not be directly
obtained by the integration of the corresponding energy density. In this
section, to consider the total vacuum energy for the configuration of two
concentric spheres, we follow the procedure which have been frequently used
in the calculations of the Casimir energy for various boundary geometries
and is based on the zeta function technique (see, for instance, \cite%
{Eliz94,Kirs02} and references therein). In the region between the spheres
the total vacuum energy is the sum of zero-point energies of elementary
oscillators:
\begin{equation}
E_{a\leq r\leq b}=-\sum_{j=1/2}^{\infty }(2j+1)\sum_{\sigma
=0,1}\sum_{s=1}^{\infty }(\gamma _{\nu _{\sigma },s}^{2}/a^{2}+M^{2})^{1/2}.
\label{toten1}
\end{equation}%
To regularize the divergent expression on the right of this formula we
introduce the related zeta function%
\begin{equation}
\zeta (u)=\mu ^{u+1}\sum_{j=1/2}^{\infty }(2j+1)\zeta _{j}\left( u\right) ,
\label{zeta}
\end{equation}%
where the parameter $\mu $ with dimension of mass is introduced by
dimensional reasons, and the partial zeta function is defined as%
\begin{equation}
\zeta _{j}\left( u\right) =\sum_{\sigma =0,1}\sum_{s=1}^{\infty }(\gamma
_{\nu _{\sigma },s}^{2}/a^{2}+M^{2})^{-u/2}.  \label{zetan}
\end{equation}%
The computation of the Casimir energy requires the analytic continuation of
the zeta function to the value $u=-1$. The starting point of our
consideration is the representation of the partial zeta function as a
contour integral in the complex plane $z$:%
\begin{equation}
\zeta _{j}\left( u\right) =\frac{1}{2\pi i}\sum_{\sigma
=0,1}\int_{C}dz\,(z^{2}+M^{2})^{-u/2}\frac{\partial }{\partial z}\ln \left[
z^{-2}C_{\nu _{\sigma }}(\eta ,az)\right] ,  \label{zetanint1}
\end{equation}%
where $C_{\nu _{\sigma }}(\eta ,az)$ is given by (\ref{Cnu}) and
$C$ denotes a closed counterclockwise contour enclosing all zeros
$\gamma _{\nu _{\sigma },s}$. The additional factor in the
argument of the log function is introduced to cancel the pole at
$z=0$. We assume that the contour $C$ is made of a large
semicircle (with radius tending to infinity) centered at the
origin and placed to its right, plus a straight part overlapping
the imaginary axis and avoiding the points $\pm iM$ by small
semicircles in the right half-plane. When the radius of the large
semicircle tends to infinity the corresponding contribution into
$\zeta _{j}\left( u\right) $ vanishes for $\mathrm{Re}\,u>1$. The
integral on the
right of Eq. (\ref{zetanint1}) can be presented in the form%
\begin{eqnarray}
\zeta _{j}\left( u\right)  &=&\zeta _{j}^{\mathrm{(int)}}\left( u\right)
+\zeta _{j}^{\mathrm{(ext)}}\left( u\right)   \notag \\
&&+\frac{1}{2\pi i}\sum_{\sigma
=0,1}\sum_{n=1,2}\int_{C^{n}}dz\,(z^{2}+M^{2})^{-\frac{u}{2}}\frac{\partial
}{\partial z}\ln \left[ 1-\frac{\tilde{J}_{\nu _{\sigma }}^{(a)}(az)\tilde{H}%
_{\nu _{\sigma }}^{(nb)}(bz)}{\tilde{H}_{\nu _{\sigma }}^{(na)}(az)\tilde{J}%
_{\nu _{\sigma }}^{(b)}(bz)}\right] ,  \label{zetanint2}
\end{eqnarray}%
where $C^{1}$ and $C^{2}$ are the upper and lower halves of the contour $C$,
$H_{\nu }^{(1,2)}(z)$ are the Hankel functions and we have introduced the
notations%
\begin{eqnarray}
\zeta _{j}^{\mathrm{(int)}}\left( u\right)  &=&\frac{1}{2\pi i}\sum_{\sigma
=0,1}\int_{C}dz\,(z^{2}+M^{2})^{-\frac{u}{2}}\frac{\partial }{\partial z}\ln %
\left[ z^{-\nu _{\sigma }-n_{\sigma }-1}\tilde{J}_{\nu _{\sigma }}^{(b)}(bz)%
\right] ,  \label{zetajint} \\
\zeta _{j}^{\mathrm{(ext)}}\left( u\right)  &=&\frac{1}{2\pi i}\sum_{\sigma
=0,1}\sum_{n=1,2}\int_{C^{n}}dz\,(z^{2}+M^{2})^{-\frac{u}{2}}\frac{\partial
}{\partial z}\ln \left[ z^{\nu _{\sigma }+n_{\sigma }-1}\tilde{H}_{\nu
_{\sigma }}^{(na)}(az)\right] .  \label{zetajext}
\end{eqnarray}%
As before, the additional factors in the arguments of the log functions in (%
\ref{zetajint}) and (\ref{zetajext}) are introduced in order to cancel the
poles at the origin. First of all let us consider the function (\ref%
{zetajint}). Noting that the eigenmodes in the region inside a
single spherical shell with radius $b$ are zeros of the function
$\tilde{J}_{\nu _{\sigma }}^{(b)}(bz)$, we see that $\zeta
_{j}^{\mathrm{(int)}}\left( u\right) $ is the partial zeta
function for this region. It can be further simplified by
making use of the relation%
\begin{eqnarray}
\prod_{\sigma =0,1}z^{-\nu _{\sigma }-n_{\sigma }-1}\tilde{J}_{\nu
_{\sigma }}^{(b)}(bz) &=&b^{2}z^{-2\nu }\left( \sqrt{%
z^{2}+M^{2}}-M\right) \nonumber \\
&& \times \left[ J_{\nu }^{2}(bz)-\frac{2M}{z}J_{\nu }(bz)J_{\nu
-1}(bz)-J_{\nu -1}^{2}(bz)\right] , \label{relln1}
\end{eqnarray}%
where $\nu $ is defined by formula (\ref{notnunu1}) with $l=j+1/2$. The
factor $(\sqrt{z^{2}+M^{2}}-M)/z^{2}$ on the right of this relation has no
zeros and singularities inside the contour $C$ and, hence, does not give
contribution into the zeta function. Now by making use the standard
properties of the Bessel functions, we see that the parts of the integrals
over $(0,\pm iM)$ cancel and we find the following integral representation
for the interior zeta function%
\begin{eqnarray}
\zeta ^{\mathrm{(int)}}(u) &=&\frac{2}{\pi }\mu ^{u+1}\sin \frac{\pi u}{2}%
\sum_{l=1}^{\infty }l\int_{M}^{\infty }dx\,(x^{2}-M^{2})^{-\frac{u}{2}}\frac{%
\partial }{\partial x}\ln \Bigg\{x^{-2\nu +2}  \notag \\
&&\times \left[ I_{\nu }^{\prime 2}(bx)+2\frac{Mb+\nu }{bx}I_{\nu
}(bx)I_{\nu }^{\prime }(bx)+\left( 1+\frac{2M\nu }{bx^{2}}+\frac{\nu ^{2}}{%
b^{2}x^{2}}\right) I_{\nu }^{2}(bx)\right] \Bigg\}.  \label{zetaint}
\end{eqnarray}%
By the same way it can be seen that the zeta function corresponding to the
part (\ref{zetajext}) is presented in the form%
\begin{eqnarray}
\zeta ^{\mathrm{(ext)}}(u) &=&\frac{2}{\pi }\mu ^{u+1}\sin \frac{\pi u}{2}%
\sum_{l=1}^{\infty }l\int_{M}^{\infty }dx\,(x^{2}-M^{2})^{-\frac{u}{2}}\frac{%
\partial }{\partial x}\ln \Bigg\{x^{-2\nu +2}  \notag \\
&&\times \left[ K_{\nu }^{\prime 2}(ax)+2\frac{Ma+\nu }{ax}K_{\nu
}(ax)K_{\nu }^{\prime }(ax)+\left( 1+\frac{2M\nu }{ax^{2}}+\frac{\nu ^{2}}{%
a^{2}x^{2}}\right) K_{\nu }^{2}(ax)\right] \Bigg\}.  \label{zetaext}
\end{eqnarray}%
Now the complete zeta function can be written in the form%
\begin{eqnarray}
\zeta \left( u\right)  &=&\zeta ^{\mathrm{(int)}}\left( u\right) +\zeta ^{%
\mathrm{(ext)}}\left( u\right) +\frac{2}{\pi }\mu ^{u+1}\sin \frac{\pi u}{2}%
\sum_{l=1}^{\infty }l  \notag \\
&&\times \sum_{\beta =\pm }\int_{M}^{\infty }dx\,(x^{2}-M^{2})^{-\frac{u}{2}}%
\frac{\partial }{\partial x}\ln \left[ 1-\frac{I_{\nu }^{(a\beta
)}(ax)K_{\nu }^{(b\beta )}(bx)}{K_{\nu }^{(a\beta )}(ax)I_{\nu }^{(b\beta
)}(bx)}\right] ,  \label{zetau}
\end{eqnarray}%
where we have used relation (\ref{relsig01}) to see that different
parities corresponding to $\sigma =0,1$ give the same contribution
into the interference part. The term $\zeta
^{\mathrm{(ext)}}\left( u\right) $ on the right of this formula is
the zeta function for the region outside a single sphere with
radius $a$. This can be seen by various ways. First of all in the
limit $b\to  \infty $ we expect that the interference effects
should disappear and the zeta function should be a sum of the zeta
functions for single spheres. As we consider the region
$a\leqslant r\leqslant b$ this sum should involve the interior
zeta function for the sphere with radius $b$ and the
exterior zeta function for the sphere with radius $a$. Indeed, in the limit $%
b\to  \infty $ the last term on the right of formula (\ref{zetau})
vanishes and the zeta function is the sum of single sphere parts.
Another way to see that $\zeta ^{\mathrm{(ext)}}\left( u\right) $
is the zeta function for the exterior region of a single sphere is
to use the method based on the scattering theory and described in
Refs. \cite{Bord01,Kirs02}.
We could also directly write down the expression for $\zeta ^{\mathrm{(ext)}%
}\left( u\right) $ by using the property that the quantities (both local and
global) for the exterior region of a single sphere are obtained from those
for the interior region by the replacements $I_{\nu }\rightleftarrows K_{\nu
}$ in the corresponding formulae. It can be easily seen that for the special
case $\alpha =1$ formulae (\ref{zetaint}) and (\ref{zetaext}) coincide with
the integral representations of the zeta functions on the Minkowski bulk
derived in Ref. \cite{Eliz98}. The last term on the right of formula (\ref%
{zetau}) is finite at $u=-1$ and the further analytic continuation
is necessary for the single sphere zeta functions only. The
corresponding procedure is standard and has been multiply used in
the calculations of the Casimir energy (see \cite{Bord01} and
references therein). For the case of a massive spinor field in the
Minkowski bulk this is demonstrated in \cite{Eliz98}. Note that
already in this special case the calculations are lengthy. The
corresponding results for the global monopole bulk depend on
additional parameter $\alpha $ and will be presented in our
forthcoming paper. Here we will be concentrated on the
interference part of the vacuum energy.

On the base of the representation for the zeta function, the total
vacuum energy in
the region $a\leqslant r\leqslant b$ is presented in the form%
\begin{equation}
E_{a\leqslant r\leqslant b}=\zeta (u)|_{u=-1}=E_{r\geqslant
a}^{(a)}+E_{r\leqslant b}^{(b)}+\Delta E,  \label{Eab}
\end{equation}%
where $E_{r\geqslant a}^{(a)}$ ($E_{r\leqslant b}^{(b)}$) is the
vacuum energy for the region outside (inside) a spherical shell
with radius $a$ ($b$) and the
interference term is given by the formula%
\begin{eqnarray}
\Delta E &=&\frac{2}{\pi }\sum_{l=1}^{\infty }l\sum_{\beta =\pm
}\int_{M}^{\infty }dx\,\sqrt{x^{2}-M^{2}}\frac{\partial }{\partial x}\ln %
\left[ 1-\frac{I_{\nu }^{(a\beta )}(ax)K_{\nu }^{(b\beta )}(bx)}{K_{\nu
}^{(a\beta )}(ax)I_{\nu }^{(b\beta )}(bx)}\right]   \notag \\
&=&-\frac{4}{\pi }\sum_{l=1}^{\infty }l\int_{M}^{\infty }dx\frac{x}{\sqrt{%
x^{2}-M^{2}}}\ln \left\vert 1-\frac{I_{\nu }^{(a+)}(ax)K_{\nu }^{(b+)}(bx)}{%
K_{\nu }^{(a+)}(ax)I_{\nu }^{(b+)}(bx)}\right\vert ,  \label{DeltaE}
\end{eqnarray}%
where we have taken into account that $F^{(w-)}(wx)=F^{(w+)\ast
}(wx)$. Now comparing with formula (\ref{pwint1}) we see that the
interaction energy and
the interaction forces are related by the formula%
\begin{equation}
p_{\mathrm{(int)}}^{(w)}=-\frac{n^{(w)}}{4\pi w^{2}}\frac{\partial }{%
\partial w}\Delta E.  \label{pwintDeltaE}
\end{equation}%
which corresponds to the standard relation $p=-\partial E/\partial V$, with $%
V$ being the volume. By using the relations%
\begin{equation}
\frac{I_{\nu }^{(a+)}(ax)K_{\nu }^{(b+)}(bx)}{K_{\nu }^{(a+)}(ax)I_{\nu
}^{(b+)}(bx)}=-\frac{\left[ xI_{\nu -1}(ax)/I_{\nu }(ax)-B^{(+)}\right] %
\left[ xK_{\nu }(bx)/K_{\nu -1}(bx)-B^{(-)}\right] }{\left[ xK_{\nu
}(ax)/K_{\nu -1}(ax)+B^{(-)}\right] \left[ xI_{\nu -1}(bx)/I_{\nu
}(bx)+B^{(+)}\right] },  \label{relebac}
\end{equation}%
and the inequalities $I_{\nu -1}(x)>I_{\nu }(x)$, $K_{\nu }(x)>K_{\nu -1}(x)$%
, $I_{\nu -1}(x)K_{\nu -1}(x)>I_{\nu }(x)K_{\nu }(x)$ for the Bessel
modified functions, it can be seen that the integrand in Eq. (\ref{DeltaE})
is positive and, hence, the interaction part of the vacuum energy is
negative. Moreover, the integrand is an increasing function with respect to $%
a$ and a decreasing function with respect to $b$. In accordance with formula
(\ref{pwintDeltaE}) this leads to negative $p_{\mathrm{(int)}}^{(w)}$
corresponding to attractive interaction forces between the spheres. To
obtain the total Casimir energy for the configuration of two spheres, $E$,
we need to add to the energy in the region between the spheres, given by Eq.
(\ref{DeltaE}), the energies coming from the regions $r\leqslant a$ and $r\geqslant b$%
. As a result one receives%
\begin{equation}
E=E^{(a)}+E^{(b)}+\Delta E,  \label{Etot}
\end{equation}%
where $E^{(w)}$ is the Casimir energy for a single spherical shell
with radius $r=w$ in the global monopole spacetime. This energy
can be evaluated by using the integral representations of the zeta
functions for the regions inside and outside of a single sphere
given by formulae (\ref{zetaint}) and (\ref{zetaext}). As it has
been mentioned above, the corresponding procedure is similar to
that previously employed in Refs. \cite{Eliz98,Cogn01} for the
calculation of the fermionic Casimir energy for a spherical shell
in the Minkowski background and will be presented in the
forthcoming paper.

Formula (\ref{DeltaE}) for the interaction part of the vacuum energy can
also be obtained by the integration of the interference part of the energy
density over the region between the spheres. Noting that the functions $%
G_{\nu }^{(w\beta )}(wx,rx)$ and $G_{\nu -1}^{(w\beta )}(wx,rx)$ are
modified cylindrical functions with respect to the argument $r$, we can
evaluate the corresponding integrals on the base of standard formulae \cite%
{Prud86} for the integrals involving the square of a modified cylindrical
function. By making use of relations (\ref{Gnuxx}) and additionally%
\begin{eqnarray}
\frac{\partial }{\partial y}G_{\nu }^{(w\beta )}(wx,y)\big|_{y=wx} &=&\frac{%
n^{(w)}wB^{(\beta )}+\nu }{wx}, \\
\frac{\partial }{\partial y}G_{\nu -1}^{(w\beta )}(wx,y)\big|_{y=wx} &=&%
\frac{\nu -1}{wx^{2}}n^{(w)}B^{(\beta )}-1,  \label{Gnxxder}
\end{eqnarray}%
one finds%
\begin{equation}
4\pi \alpha ^{2}\int_{a}^{b}dr\,r^{2}\varepsilon ^{(ab)}(r)=E_{b\leqslant
r<\infty }^{(a)}+E_{0\leqslant r\leqslant a}^{(b)}+\Delta E,
\label{integinterf}
\end{equation}%
with $\Delta E$ defined by formula (\ref{DeltaE}). In (\ref{integinterf}), $%
E_{b\leqslant r<\infty }^{(a)}$ ($E_{0\leqslant r\leqslant a}^{(b)}$) is the
vacuum energy for a single spherical shell with radius $a$ ($b$) in the
region $b\leqslant r<\infty $ ($0\leqslant r\leqslant a$). In deriving (\ref%
{integinterf}), we have used the formula for $\varepsilon ^{(ab)}(r)$ in
terms of $\varepsilon _{ab}(r)$ from (\ref{qabrl}) at the lower limit of
integration and in terms of $\varepsilon _{ba}(r)$ from (\ref{qbar}) at the
upper limit. Now adding to the part (\ref{integinterf}) the parts coming
from single spheres, corresponding to the term $q_{w}(r)$ in (\ref{qdec}),
for the boundary part of the vacuum energy in the region $a\leqslant
r\leqslant b$ we obtain the formula (\ref{Eab}). Hence, we have explicitly
checked that the interaction part of the vacuum energy evaluated from the
sum of the zero-point energies of elementary oscillators coincides with the
corresponding energy obtained by the integration of the local density. Note
that for a scalar field, in general, this is not the case and in the
discussion of the energy balance it is necessary to take into account the
surface part of the energy located on the boundaries (see, for instance,
Refs. \cite{Kenn80} for various special cases of boundary geometries and
Ref. \cite{Saha04} for the general case of bulk and boundary geometries). In
this case the relation (\ref{pwintDeltaE}) between the vacuum forces and the
total energy is also modified by the presence of an additional term coming
form the surface energy.

The general formula for the interaction part of the vacuum energy is
simplified in limiting cases. First we consider the limit when $%
a,b\to  \infty $ and $b-a$ is fixed. In this limit the main
contribution comes from large values $l$ and we can use the
uniform asymptotic expansions for the Bessel modified functions
for large values of the order (see, for instance, \cite{Abra64}).
Replacing the Bessel modified functions by the leading terms of
these asymptotic expansions, we introduce a new integration
variable $u=\sqrt{z^{2}-M^{2}}$ and replace the summation over $l$
by the integration: $\sum_{l=1}^{\infty }f(l)\to  \alpha
\int_{0}^{\infty }d\nu f(\alpha \nu )$. Further, introducing polar
coordinates in the $(\nu ,u)$ plane, after the integration of the
angular
part one obtains the following result%
\begin{equation}
\frac{\Delta E}{4\pi \alpha ^{2}a^{2}}\approx -\frac{1}{\pi ^{2}}%
\int_{M}^{\infty }dx\,x\sqrt{x^{2}-M^{2}}\ln \left[ 1+\frac{x-M}{x+M}%
e^{-2(b-a)x}\right] .  \label{limit1DeltaE}
\end{equation}%
The expression on the right coincides with the interference part of the
Casimir energy per unit surface for parallel plates with bag boundary
conditions in the Minkowski bulk (see, for instance, \cite{Grib94}).

In the limit $a\to  0$ the main contribution to the energy comes
from $l=1$ term. By using the formulae for the Bessel modified
functions for
small values of the argument in the leading order one finds%
\begin{eqnarray}
\Delta E &\approx &-\frac{8M}{\pi \Gamma ^{2}(\nu )}\left( \frac{a}{2b}%
\right) ^{\frac{2}{\alpha }}\int_{\mu _{b}}^{\infty }dx\frac{x^{\frac{2}{%
\alpha }}}{\sqrt{x^{2}-\mu _{b}^{2}}}  \label{DeltaEsmalla} \\
&&\times \frac{\bar{I}_{\nu }^{(b)}(x)\bar{K}_{\nu }^{(b)}(x)+(x^{2}-\mu
_{b}^{2})I_{\nu }(x)K_{\nu }(x)+(x^{2}-\mu _{b}^{2})/\mu _{b}}{\bar{I}_{\nu
}^{(b)2}(x)+(x^{2}-\mu _{b}^{2})I_{\nu }^{2}(x)},  \notag
\end{eqnarray}%
where $\nu =1/\alpha +1/2$ and for a function $F(z)$ we have introduced
notation%
\begin{equation}
\bar{F}^{(w)}(x)=xF^{\prime }(x)+(n^{(w)}\mu _{w}+\nu )F(x).
\label{Fbarbnot}
\end{equation}%
This formula is further simplified in the case of a massless spinor field:%
\begin{equation}
\Delta E\approx -\frac{8}{\pi b\Gamma ^{2}(\nu )}\left( \frac{a}{2b}\right)
^{\frac{2}{\alpha }}\int_{0}^{\infty }dx\frac{x^{\frac{2}{\alpha }-1}}{%
I_{\nu -1}^{2}(x)+I_{\nu }^{2}(x)}.  \label{DeltaEsmallaM0}
\end{equation}%
In the limit $b\to  \infty $ assuming that $Mb\gg 1$, the main
contribution into the integral over $x$ in (\ref{DeltaE}) comes
from the
lower limit and to the leading order we have the formula%
\begin{equation}
\Delta E\approx -\sqrt{\frac{\pi }{Mb}}\frac{e^{-2Mb}}{b\alpha }%
\sum_{l=1}^{\infty }l^{2}\frac{I_{\nu -1}(Ma)-I_{\nu }(Ma)}{K_{\nu
-1}(Ma)+K_{\nu }(Ma)},  \label{DeltaElargeb}
\end{equation}%
with the exponentially suppressed interference part. In the case of a
massless spinor field and for large values $b$ the asymptotic behavior of
the interference part in the vacuum energy is given by formula (\ref%
{DeltaEsmallaM0}).

\begin{figure}[tbph]
\begin{center}
\epsfig{figure=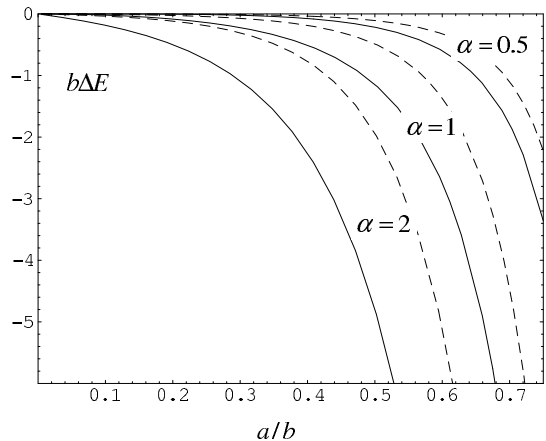,width=7.5cm,height=6.cm}
\end{center}
\caption{Interaction part of the vacuum energy as a function of $a/b$ for
various values of the parameter $\protect\alpha $. The full curves
correspond to the massless fermionic field and the dashed curves are for the
massive field with $Mb=1$.}
\label{fig1}
\end{figure}

Now let us consider the interaction part of the vacuum energy in the
limiting cases for the parameter $\alpha $. The limit $\alpha \ll 1$
corresponds to strong gravitational fields. In accordance with formula (\ref%
{notnunu1}), in this limit the order of the Bessel modified functions is
large. By using the uniform asymptotic expansions for these functions, we
can see that the main contribution comes from $l=1$ term and to the leading
order one finds%
\begin{equation}
\Delta E\approx -\frac{2\sqrt{2}}{b}\frac{(a/b)^{2/\alpha }}{\sqrt{\pi
\alpha (1-a^{2}/b^{2})}}.  \label{DeltaEsmallalfa}
\end{equation}%
In the same limit, to the leading order for the interaction forces between
the spheres one has $p_{\mathrm{(int)}}^{(w)}\approx \Delta E/(2\pi \alpha
w^{3})$. In the limit $\alpha \gg 1$ corresponding to negative solid angle
deficit the main contribution into the interaction part of the vacuum energy
comes from large values $l$. By the way similar to that used in section \ref%
{sec:vevemt} for the vacuum densities, it can be seen that to the leading
order over $\alpha $ one has%
\begin{equation}
\Delta E\approx -\frac{4\alpha ^{2}}{\pi }\int_{1/2}^{\infty
}dy\int_{M}^{\infty }dx\,\frac{x(y-1/2)}{\sqrt{x^{2}-M^{2}}}\ln \left\vert 1-%
\frac{I_{y}^{(a+)}(ax)K_{y}^{(b+)}(bx)}{K_{y}^{(a+)}(ax)I_{y}^{(b+)}(bx)}%
\right\vert .  \label{DeltaElargealf}
\end{equation}%
It follows from here that the interaction energy per unit surface tends to
the finite value. In figure \ref{fig1} we have plotted the interaction part
of the vacuum energy as a function on the ratio of the sphere radii for
different values of the parameter $\alpha $ determining the solid angle
deficit. The full curves correspond to the massless fermionic field and the
dashed curves are for the massive field with $Mb=1$. The curves for $\alpha
=1$ correspond to the bulk geometry of Minkowski spacetime.

\section{Conclusion}

\label{sec:conc}

In this paper we have studied the fermionic Casimir effect for the geometry
of two spherical shells in a idealized point-like global monopole spacetime.
The VEVs of the energy-momentum tensor and the fermionic condensate are
investigated in the region between two spheres assuming that on the sphere
surfaces the MIT bag boundary condition is satisfied. The unrenormalized
expectation values are expressed as series over the zeros of a combination
of the cylindrical functions given by formula (\ref{Cnu}) with the tilted
notation from (\ref{tildenot}). To deal with this type of series, in
Appendix \ref{sec:app1} we derive summation formula by making use of the
generalized Abel-Plana formula. The application of this formula to the
corresponding mode-sums allowed us to extract from the VEVs the parts due to
single spheres when the second sphere is absent and to present the
additional part in terms of exponentially convergent integrals. In
particular, the latter are convenient for numerical calculations. We have
shown that the different parities for the fermionic eigenfunctions give the
same contribution into the VEVs. For the points away from the boundaries,
the boundary induced terms are unamiguously defined and the ambiguities in
the renormalization procedure in the form of an arbitrary mass scale are
contained in the boundary-free parts only. In particular, for a massless
fermionic field the boundary induced energy-momentum tensor does not contain
conformal anomalies and is traceless. We have discussed the behavior of the
VEVs for the energy-momentum tensor and fermionic condensate in various
limiting cases. In particular, for strong gravitational fields corresponding
to small values of the parameter $\alpha $, these VEVs are exponentially
suppressed. For large values $\alpha $ corresponding to the negative solid
angle deficit, $\alpha \gg 1$, the VEVs tend to the finite limiting value.
By using the radial vacuum stress, in section \ref{sec:intforce} we consider
the vacuum forces acting on the spheres. These forces are presented in the
form of the sum of self-action and interaction terms. Due to the surface
divergences in the VEVs of the energy-momentum tensor, the self-action part
of the vacuum force is divergent and needs further renormalization. This can
be done by making use of zeta function renormalization technique. The
interaction parts in the vacuum forces are finite for all non-zero distances
between the spheres and are given by formula (\ref{pwint}). An alternative
representation for these forces is given by Eq. (\ref{pwint1}). The Casimir
energy in the region between two spheres is considered in section \ref%
{sec:Casen}. By using the zeta function technique we have presented it as
the sum of the single sphere and interaction energies. The latter is given
by formula (\ref{DeltaE}) and is negative. We have explicitly checked that
the interaction parts in the vacuum forces and in the vacuum energy are
related by standard formula (\ref{pwintDeltaE}). In addition, we have shown
that the interaction part of the vacuum energy evaluated from the sum of the
zero-point energies of elementary oscillators coincides with the
corresponding energy obtained by the integration of the local density. Note
that for a scalar field, in general, this is not the case and in the
discussion of the energy balance it is necessary to take into account the
surface part of the energy located on the boundaries. We have investigated
the interaction part of the vacuum energy in various limiting cases. First,
we have checked that in the limit $a,b\to  \infty $, with fixed $b-a$%
, the result for the geometry of parallel plates with bag boundary
conditions in the Minkowski bulk is obtained. In the limit $a\to
0$ and for fixed values for the other parameters the interference
part of the vacuum energy vanishes as $a^{2/\alpha }$. For large
values $b$ this energy behaves like $b^{-2/\alpha -1}$ for a
massless field and is exponentially suppressed for a massive field
under the condition $Mb\gg 1$. For small values of the parameter
$\alpha $ corresponding to strong gravitational fields the vacuum
energy is suppressed by the factor $(a/b)^{2/\alpha }$. For large
values $\alpha $ corresponding to a negative solid angle deficit,
the interaction part of the vacuum energy per unit surface tends
to the finite value given by formula (\ref{DeltaElargealf}).

\section*{Acknowledgement}

AAS was supported by PVE/CAPES program and in part by the Armenian Ministry
of Education and Science Grant No. 0124. ERBM thanks Conselho Nacional de
Desenvolvimento Cient\'\i fico e Tecnol\'ogico (CNPq.) and FAPESQ-PB/CNPq.
(PRONEX) for partial financial support.

\appendix

\section{Appendix: Summation formulae over zeros of $C_{\protect\nu }(%
\protect\eta ,z)$}

\label{sec:app1}

As we have seen in section \ref{sec:vevemt}, for the massive spinor field in
the region between two spheres the eigenmodes for the quantum number $k$ are
expressed in terms of the zeros of the function
\begin{equation}
C_{\nu }(\eta ,z)\equiv \tilde{J}_{\nu }^{(a)}(z)\tilde{Y}_{\nu }^{(b)}(\eta
z)-\tilde{Y}_{\nu }^{(a)}(z)\tilde{J}_{\nu }^{(b)}(\eta z),  \label{bescomb1}
\end{equation}%
where the tilded quantities are defined as (\ref{tildenot}) and we
will assume that $\eta >1$. To evaluate the VEV of the
energy-momentum tensor we need to sum series over these zeros.
Here we derive a summation formula for this type of series by
making use of the generalized Abel-Plana formula
derived in \cite{Sahmat, Sahrev}. In this formula as functions $g(z)$ and $%
f(z)$ we choose
\begin{equation}
g(z)=\frac{1}{2i}\left[ \frac{\tilde{H}_{\nu }^{(1b)}(\eta z)}{\tilde{H}%
_{\nu }^{(1a)}(z)}+\frac{\tilde{H}_{\nu }^{(2b)}(\eta z)}{\tilde{H}_{\nu
}^{(2a)}(z)}\right] \frac{h(z)}{C_{\nu }(\eta ,z)},\quad f(z)=\frac{h(z)}{%
\tilde{H}_{\nu }^{(1a)}(z)\tilde{H}_{\nu }^{(2a)}(z)},  \label{gefcomb}
\end{equation}%
with the sum and difference
\begin{equation}
g(z)-(-1)^{n}f(z)=-i\frac{\tilde{H}_{\nu }^{(na)}(\eta z)}{\tilde{H}_{\nu
}^{(na)}(z)}\frac{h(z)}{C_{\nu }(\eta ,z)},\quad n=1,2.  \label{gefsumnew}
\end{equation}%
The conditions for the generalized Abel-Plana formula written in terms of
the function $h(z)$ are as follows
\begin{equation}
|h(z)|<\varepsilon _{1}(x)e^{c|y|},\quad |z|\to  \infty ,\quad
z=x+iy, \label{cond31}
\end{equation}%
where $c<2(\eta -1)$ and $\varepsilon _{1}(x)/x\to  0$ for $%
x\to  +\infty $. Let $\gamma _{\nu ,s}$ be positive zeros for the
function $C_{\nu }(\eta ,z)$. To find the residues of the function
$g(z)$ at the poles $z=\gamma _{\nu ,s}$ we need the derivative
\begin{equation}
\frac{\partial }{\partial z}C_{\nu }(\eta ,z)=\frac{4}{\pi T_{\nu
}^{ab}(\eta ,z)}\frac{\tilde{J}_{\nu }^{(b)}(\eta z)}{\tilde{J}_{\nu
}^{(a)}(z)}\,,\quad z=\gamma _{\nu ,s},  \label{CABderivative}
\end{equation}%
where we have introduced the notation
\begin{equation}
T_{\nu }^{ab}(\eta ,z)=z\left[ \frac{\tilde{J}_{\nu }^{(a)2}(z)}{\tilde{J}%
_{\nu }^{(b)2}(\eta z)}D_{\nu }^{(b)}-D_{\nu }^{(a)}\right] ^{-1},
\label{tekaAB}
\end{equation}%
with%
\begin{equation}
D_{\nu }^{(w)}=w^{2}\left[ k^{2}+(M-E)(M-n_{\sigma }\nu n^{(w)}/w)+\frac{%
n^{(w)}k^{2}}{2wE}\right] ,  \label{Dnuj}
\end{equation}%
and $ka=z$. By using this relation it can be seen that
\begin{equation}
\underset{z=\gamma _{\nu ,s}}{\mathrm{Res}}g(z)=\frac{\pi }{4i}T_{\nu
}^{ab}(\eta ,\gamma _{\nu ,s})h(\gamma _{\nu ,s}).  \label{rel31}
\end{equation}%
Let $h(z)$ be an analytic function for $\mathrm{Re\,}z\geq 0$ except
possible branch points on the imaginary axis. By using the generalized
Abel-Plana formula, similar to the derivation of summation formula (4.13) in
\cite{Sahrev}, it can be seen that if it satisfies condition (\ref{cond31}),
\begin{equation}
h(ze^{\pi i})=-h(z)+o(z^{-1}),\quad z\to  0,  \label{cor3cond1}
\end{equation}%
and the integral
\begin{equation}
\int_{0}^{L}\frac{h(x)dx}{\tilde{J}_{\nu }^{(a)2}(x)+\tilde{Y}_{\nu
}^{(a)2}(x)}  \label{cor2cond2}
\end{equation}%
exists, then
\begin{eqnarray}
&&\lim_{L\to  +\infty }\left\{ \frac{\pi ^{2}}{4}\sum_{s=1}^{m}h(%
\gamma _{\nu ,s})T_{\nu }^{ab}(\eta ,\gamma _{\nu ,s})-\int_{0}^{L}\frac{%
h(x)dx}{\tilde{J}_{\nu }^{(a)2}(x)+\tilde{Y}_{\nu }^{(a)2}(x)}\right\} =
\notag \\
&=&-\frac{\pi }{2}\underset{z=0}{\mathrm{Res}}\left[ \frac{h(z)\tilde{H}%
_{\nu }^{(1b)}(\eta z)}{C_{\nu }(\eta ,z)\tilde{H}_{\nu }^{(1a)}(z)}\right] -%
\frac{\pi }{4}\sum_{\beta =\pm }\int_{0}^{\infty }dx\,\Omega _{a\nu
}^{(\beta )}(x,\eta x)h(xe^{\beta \pi i/2}),  \label{cor3form}
\end{eqnarray}%
where $\gamma _{\nu ,m}<L<\gamma _{\nu ,m+1}$. Here the function $\Omega
_{a\nu }^{(\beta )}(x,\eta x)$ is defined as%
\begin{equation}
\Omega _{a\nu }^{(\beta )}(x,\eta x)=\frac{K_{\nu }^{(b\beta )}(\eta
x)/K_{\nu }^{(a\beta )}(x)}{K_{\nu }^{(a\beta )}(x)I_{\nu }^{(b\beta )}(\eta
x)-I_{\nu }^{(a\beta )}(x)K_{\nu }^{(b\beta )}(\eta x)},  \label{Oma}
\end{equation}%
and for a given function $F(z)$\ we use the notation
\begin{equation}
F^{(w\pm )}(z)=zF^{\prime }(z)+\left[ n^{(w)}\left( \mu _{w}-\sqrt{%
z^{2}e^{\pm \pi i}+\mu _{w}^{2}}\right) -n_{\sigma }\nu \right] F(z)\ .
\label{Fbarpm}
\end{equation}%
In section \ref{sec:vevemt}, formula (\ref{cor3form}) is used to derive the
VEV of the energy momentum-tensor for the region between two spherical
shells in the global monopole spacetime. As it can be seen from expressions (%
\ref{fnueps})--(\ref{fnupperp}), the corresponding functions $h(z)$ satisfy
the relation
\begin{equation}
h(xe^{\pi i/2})=-h(xe^{-\pi i/2})\ ,\quad \mathrm{for}\quad 0\leqslant
x\leqslant \mu _{a}\ .  \label{relforf}
\end{equation}%
By taking into account that for these values $x$ one has $%
F^{(w+)}(wx/a)=F^{(w-)}(wx/a)$, we conclude that in this case the part of
the integral on the right of Eq. (\ref{cor3form}) over the interval $(0,\mu
_{a})$ vanishes.


\begin{thebibliography}{99}
\bibitem{Kibble} T. W. Kibble, J. Phys. A\textbf{\ 9}, 1387 (1976).

\bibitem{Vilenkin} A. Vilenkin, Phys. Rep. \textbf{121}, 263 (1985).

\bibitem{Soko77} D.D. Sokolov and A.A. Starobinsky, Dokl. Akad. Nauk USSR
\textbf{234}, 1043 (1977).

\bibitem{B-V} M. Barriola and A. Vilenkin, Phys. Rev. Lett. \textbf{63}, 341
(1989).

\bibitem{Volo98} G.E. Volovik, Pisma Zh. Eksp. Teor. Fiz. \textbf{67}, 666
(1998) [JETP Lett. \textbf{67}, 698 (1998)].

\bibitem{M-L} F.D. Mazzitelli and C.O. Lousto, Phys. Rev. D \textbf{43}, 468
(1991).

\bibitem{EVN} E.R. Bezerra de Mello, V.B. Bezerra, and N.R. Khusnutdinov,
Phys. Rev. D \textbf{60}, 063506 (1999).

\bibitem{MKS} M. Bordag, K. Kirsten, and S. Dowker, Commun. Math. Phys.
\textbf{128}, 371 (1996).

\bibitem{Dowk99} J.S. Dowker and K. Kirsten, Commun. Anal. Geom.
\textbf{7}, 641 (1999).

\bibitem{C-E} F.C. Carvalho and E.R. Bezerra de Mello, Class. Quantum Grav.
\textbf{18}, 1637 (2001); Class. Quantum Grav. \textbf{18}, 5455 (2001).

\bibitem{E} E.R. Bezerra de Mello, J. Math. Phys. \textbf{43}, 1018 (2002).

\bibitem{EVN1} E.R. Bezerra de Mello, V.B. Bezerra, and N.R. Khusnutdinov,
J. Math. Phys. \textbf{42}, 562 (2001).

\bibitem{A-M} A.A. Saharian and M.R. Setare, Class. Quantum Grav. \textbf{20}%
, 3765 (2003).

\bibitem{Sahmat} A.A. Saharian, Izv. Akad. Nauk. Arm. SSR. Mat. \textbf{22},
166 (1987) [Sov. J. Contemp. Math. Anal. \textbf{22}, 70 (1987)].

\bibitem{Sahrev} A.A. Saharian, "The Generalized Abel-Plana Formula.
Applications to Bessel functions and Casimir effect," Report No. IC/2000;
hep-th/0002239.

\bibitem{Saha-Mello} A.A. Saharian and E.R. Bezerra de Mello, J. Phys. A
\textbf{37}, 3543 (2004).

\bibitem{Saha03b} A.A. Saharian and M.R. Setare, Int. J. Mod. Phys. A
\textbf{19}, 4301 (2004).

\bibitem{Bend76} C.M. Bender and P. Hays, Phys. Rev. D \textbf{14}, 2622
(1976).

\bibitem{Milt80} K.A. Milton, Phys. Rev. D \textbf{22}, 1444 (1980).

\bibitem{Milt81} K.A. Milton, Phys. Lett. B \textbf{104}, 49 (1981).

\bibitem{Milt83} K.A. Milton, Ann. Phys. (N.Y.) \textbf{150}, 432 (1983).

\bibitem{Baac83} J. Baacke and Y. Igarashi, Phys. Rev. D \textbf{27}, 460
(1983).

\bibitem{Blau88} S.K. Blau, M. Wisser, and A. Wipf, Nucl. Phys. B\textbf{\
310}, 163 (1988).

\bibitem{Eliz98} E. Elizalde, M. Bordag, and K. Kirsten, J. Phys. A \textbf{%
31}, 1743 (1998).

\bibitem{Cogn01} G. Cognola, E. Elizalde, and K. Kirsten, J. Phys. A \textbf{%
34}, 7311 (2001).

\bibitem{Grib94} A.A. Grib, S.G. Mamayev, and V.M. Mostepanenko, \textit{%
Vacuum Quantum Effects in Strong Fields} (Friedmann Laboratory Publishing,
St. Petersburg, 1994).

\bibitem{Most97} V.M. Mostepanenko and N.N. Trunov, \textit{The Casimir
Effect and Its Applications} (Oxford University Press, Oxford, 1997).

\bibitem{Plun86} G. Plunien, B. Muller and W. Greiner, Phys. Rep. \textbf{134%
}, 87 (1986).

\bibitem{Milt02} K.A. Milton, \textit{The Casimir Effect: Physical
Manifestation of Zero--Point Energy} (World Scientific, Singapore, 2002).

\bibitem{Bord01} M. Bordag, U. Mohidden, and V.M. Mostepanenko, Phys. Rep.
\textbf{353}, 1 (2001).

\bibitem{Birr82} N.D. Birrell and P.C.W. Davies, \textit{Quantum Fields in
Curved Space} (Cambridge University Press, Cambridge, England, 1982).

\bibitem{Berest} V.B. Berestetskii, E.M. Lifshits, and L.~P. Pitaevskii,
\textit{Quantum Electrodynamics} (Pergamon Press, Oxford, 1982).

\bibitem{Prud86} A.P. Prudnikov, Yu.A. Brychkov, and O.I. Marichev, \textit{%
Integrals and Series} (Gordon and Breach, New York, 1986), Vol.2.

\bibitem{Eliz94} E. Elizalde, S.D. Odintsov, A. Romeo, A.A. Bytsenko, and S.
Zerbini, \textit{Zeta Regularization Techniques with Applications} (World
Scientific, Singapore, 1994).

\bibitem{Kirs02} K. Kirsten, \textit{Spectral Functions in Mathematics and
Physics} (Chapman and Hall/CRC, Boca Raton, 2002).

\bibitem{Kenn80} G. Kennedy, R. Critchley, and J.S. Dowker, Ann. Phys.
(N.Y.) \textbf{125}, 346 (1980); A. Romeo and A.A. Saharian, J. Phys. A
\textbf{35}, 1297 (2002); A.A. Saharian, Phys. Rev. D \textbf{63}, 125007
(2001); S.A. Fulling, J. Phys. A \textbf{36}, 6857 (2003); A.A. Saharian and
M.R. Setare, Class. Quantum Grav. \textbf{21}, 5261 (2004); A.A. Saharian,
Phys. Rev. D \textbf{70}, 064026 (2004); I. Cavero-Pel\'{a}ez, K.A. Milton,
and J. Wagner, hep-th/0508001; A.A. Saharian and A.S. Tarloyan,
hep-th/0603144.

\bibitem{Saha04} A.A. Saharian, Phys. Rev. D \textbf{69}, 085005 (2004).

\bibitem{Abra64} M. Abramowitz and I.A. Stegun, \textit{Handbook of
Mathematical Functions} (National Bureau of Standards, Washington DC, 1964).
\end{thebibliography}
\end{document}